\documentclass[a4paper,fleqn,usenatbib]{mnras}

\usepackage{newtxtext,newtxmath}

\usepackage[T1]{fontenc}
\usepackage{ae,aecompl}

\usepackage{graphicx}	
\usepackage{amsmath}	
\usepackage{amssymb}	

\title[Radio study of the extended TeV source VER J1907$+$062]{Radio study of the extended TeV source VER J1907$+$062}

\author[L. Duvidovich et al.]{
L. Duvidovich,$^{1}$\thanks{E-mail: lduvidovich@iafe.uba.ar}
A. Petriella,$^{1,2}$
and E. Giacani$^{1,3}$
\\
$^{1}$CONICET-Universidad de Buenos Aires, Instituto de Astronom\'ia y F\'isica del Espacio (IAFE), Buenos Aires, Argentina\\
$^{2}$Universidad de Buenos Aires, Ciclo B\'asico Com\'un, Buenos Aires, Argentina\\
$^{3}$Universidad de Buenos Aires, Facultad de Arquitectura, Dise\~{n}o y Urbanismo, Buenos Aires, Argentina
}

\date{Accepted XXX. Received YYY; in original form ZZZ}

\pubyear{2019}

\begin{document}
\label{firstpage}
\pagerange{\pageref{firstpage}--\pageref{lastpage}}
\maketitle

\begin{abstract}
This paper aims to provide new insights on the origin of the TeV source VER J1907$+$062 through new high-quality radio observations. 
We used the Karl G. Jansky Very Large Array (VLA) to observe the whole extension of VER J1907$+$062 at 1.5 GHz with a mosaicking technique 
and the PSR J1907$+$0602 in a single pointing at 6 GHz. These data were used together with $^{12}$CO and atomic hydrogen observations obtained from public surveys to investigate the interstellar medium in the direction of VER J1907$+$062. The new radio observations do not show any evidence of a pulsar wind nebula (PWN) driven by the pulsars present in the field and no radio counterpart to the proposed X-ray PWN powered by PSR J1907+0602 is seen in the new VLA image at 6 GHz down to a noise level 
of 10 $\mu$Jy beam$^{-1}$. 
Molecular clouds were discovered over the eastern, southern, and western borders of the radio shell of G40.5$-$0.5, suggesting an association with the SNR. We explored several scenarios for the origin of VER J1907+062. We propose as the most probable scenario one in which the TeV emission is produced by two separated $\gamma$-ray sources located at different distances: one of leptonic origin and associated with a PWN powered by PSR J1907$+$0602 at $\sim 3.2$ kpc and another of hadronic origin and produced by the interaction between G40.5$-$0.5 and the surrounding molecular gas at $\sim 8.7$ kpc.
\end{abstract}



\section{Introduction}
\label{introd}

The TeV source VER J1907$+$062 is an example of the so-called very high energy (VHE) $\gamma$-ray \textit{dark} sources, which are detected in the VHE range but do not have a clear counterpart at other energy bands. 
It was first discovered by the MILAGRO Collaboration and identified as the extended VHE source MGRO 1908$+$06 \citep{abdo07}.
\citet{aharo09} reported on follow-up observations with H.E.S.S. and discovered some hint of spatial variation of the TeV spectrum. They found that the emission peaks are slightly offset when the excess events are separated into two energy bands ($0.7-2.5$ TeV and $>2.5$ TeV), 
being the high-energy emission harder than the low energy one. 
However, the systematic errors forbid any conclusion on the separation in two sources or peaks.  
In the VHE regime, the region was also observed with VERITAS between 2007 and 2012 (designated as VER J1907$+$062), totalizing about 62 hr of useful exposure \citep{aliu14}. The new $\gamma$-ray excess map shows strong TeV emission near the location of the pulsar PSR J1907$+$0602, although the peak of the excess counts is offset to the west. As first observed by H.E.S.S., the TeV emission also extends toward the SNR G40.5$-$0.5 (see Fig. \ref{Radio_hess}). 

The $\gamma$-ray pulsar PSR J1907$+$0602 was first   
discovered by \citet{abdo09} using the Fermi-LAT telescope. It has a characteristic age of 19.5 kyr and a
spin-down luminosity of $3\times 10^{36}$ erg s$^{-1}$ \citep{abdo10}. These authors report on the detection of pulsed emission 
in the radio band using the Arecibo 305 m telescope. The time-averaged flux density at 1.4 GHz is 3.4 $\mu$Jy and 
the dispersion measure (DM) of 82.1 pc cm$^{-3}$ implies a distance of $3.2 \pm 0.6$ kpc, using the model 
for the electron distribution of \citet{cordes02}. 
The authors also analyze the emission of PSR J1907$+$0602 in the X-ray band using {\it Chandra} observations. 
They find a point source with a hard non-thermal spectrum and some excess of diffuse emission around it, which could 
originate in a compact pulsar wind nebula (PWN) powered by the pulsar.
 {\it XMM-Newton} observations also show that PSR J1907$+$0602 is a point source with marginal excess X-ray emission, interpreted as
a bow shock in front of the pulsar moving through the interstellar medium (ISM) \citep{pandel12}. 
\citet{pandel15} reported on the non detection of diffuse X-ray emission associated either with VER J1907$+$062 or with the SNR G40.5$-$0.5.

The SNR G40.5$-$0.5 presents in the radio band a non-thermal shell of  22$^{\prime}$ in diameter  and
a spectral index $\alpha \sim 0.4$ ($S_{\nu}\propto \alpha^{-\nu}$) \citep{sun11}.
In the radio continuum image at 1.4 GHz of the Very Large Array Galactic Plane Survey (VGPS, \citealt{stil06}), 
the shell appears well defined in the NE direction and more diffuse toward the SW. 
Using the $\Sigma - D$ relation, where $\Sigma$ is the surface brightness and $D$ the diameter of the SNR, \citet{downes80} estimated a distance between 5.5 and 8.5 kpc and an age between 20 and 40 kyr. 
Recently, the Pulsar Arecibo L-Band Feed Array (PALFA) project confirmed the presence of the pulsar PSR J1907$+$0631 lying
near the center of G40.5$-$0.5 \citep{lyne17} and the authors suggested an association between them.  
It is a young pulsar ($\tau_{c} \sim 11$ kyr) with a 
spin-down luminosity of $\sim 5 \times 10^{35}$ erg s$^{-1}$ and a DM of $\sim 430$ pc cm$^{-3}$, 
which translates into a distance of 7.9 kpc, using also the model for the electron distribution of \citet{cordes02}.   
The authors also report on the presence of a region of enhanced radio emission at the position
of PSR J1907$+$0631, which they interpret as a probable PWN powered by the pulsar. 

A third pulsar, PSR J1905$+$0600, is detected in the direction of VER J1907$+$062. It is a low energy 
($\dot{E}\sim 5.3 \times 10^{32}$ erg s$^{-1}$) and old ($\sim 6$ Myr) radio pulsar with a high DM of $\sim 730$ pc cm$^{-3}$. 
This translates into a distance of 18 kpc, which places the pulsar outside the solar circle \citep{hobbs04}. 

So far, the nature of VER J1907$+$062 has not been determined. 
Based on the strong TeV emission around PSR J1907$+$0602, which
in turn presents marginal evidence of being powering a PWN in X-rays, VER J1907$+$062 was classified as a TeV
PWN candidate \citep{aliu14}. The luminosity of the pulsar gives a TeV $\gamma$-ray efficiency similar to that of
TeV PWNe with pulsars of similar spin-down energy \citep{abdo09}. However, its large spatial
extent (larger than the rest of TeV PWNe of similar age) and the lack of variation of the $\gamma$-ray
spectrum across the TeV emission argue against interpreting it as solely powered by PSR
J1907$+$0602. Moreover, a two-dimensional Gaussian fails to accurately fit the VHE
morphology of the source, which indicates that VER J1907$+$062 may be the superposition of two
sources, either separated or interacting. On the other hand, based on the fact that the TeV emission partially overlaps the
SNR G40.5$-$0.5, \citet{aharo09} first suggested that the interaction between the remnant and the ISM could
contribute as an additional source of $\gamma$-rays.

In this paper we report on high-quality radio observations carried out with the Karl G. Jansky Very Large 
Array\footnote{The Very Large Array of the National Radio Astronomy Observatory is a facility of the National Science Foundation operated under cooperative agreement by Associated Universities, Inc.} (VLA) toward VER J1907$+$062 with the aim to investigate different scenarios about its nature. In particular, our purpose is to detect the radio counterpart of the proposed X-ray PWN driven by PSR J1907+0602, as well as explore the rol of the other two pulsars present in the observed field as possible sources of TeV emission. 
In addition, we analyze the molecular gas in the surroundings of the SNR G40.5$-$0.5 to determine the possibility that the remnant contributes to the TeV emission from VER J1907$+$062 via pion decay from proton-proton collisions. In this context, it is expected a correlation between the $\gamma$-ray emission and matter concentration.

\section{Observations and data reduction}

\subsection{New radio observations}

The radio continuum observations toward the whole extension of VER J1907$+$062 were performed using the VLA in its D configuration (project ID 18A-093). To cover the extension of this large VHE source ($\sim 62^\prime $), we used a mosaicking technique with 12 different pointings following an hexagonal pattern. The data were taking at L-band using the wide-band 1 GHz receiver system centered at 1.5 GHz which consists of 16 spectral windows with a bandwidth of 64 MHz each, spread into 64 channels. The source 3C 48 was used as primary flux density and bandpass calibrations, while the phase was calibrated with J1859+1259.

Also, we carried out observations toward the PSR J1907$+$0602 in a single pointing using the VLA in its D configuration in C-band. These observations were planned to look for the radio PWN around the pulsar, which presents a hint of a PWN in X-ray. 
In this case, we used the wide-band 4$-$8 GHz receiver system. We requested the 3-bit sampler with 2 sub-bands of 2 GHz, each comprising 16 spectral windows with a bandwidth of 128 MHz each, spread into 64 channels. The source 3C 48 was used as flux density and bandpass calibrator, and J1922$+$1530 as a phase calibrator. Table \ref{tab:Radio_observaciones} summarizes the observations. 

The data at both frequencies were processed through the VLA CASA Calibration Pipeline. We improved the quality of the calibrated data by applying extra flagging before imaging. Cleaned images were obtained with the TCLEAN task of CASA, using the multi-frequency deconvolution algorithm. We applied a Briggs weighting with a robustness parameter of 1.0 and 0.0 for the VER J1907$+$062 and PSR J1907$+$0602 images, respectively. At 1.5 GHz the radio image has a synthesized beam of $51^{\prime\prime}.1 \times 39^{\prime\prime}.5$ at 
a position angle of $21^\circ.9$, and a rms noise of 1 mJy beam$^{-1}$. The final image at 6 GHz has a resolution
of $10^{\prime\prime}.2 \times 8^{\prime\prime}.6$ at a position angle of $-48^\circ.5$, and a rms noise of 10 $\mu$Jy beam$^{-1}$.

\begin{table}
\centering
  \caption{Summary of VLA observations.}
    \begin{tabular}{ccc}
    \hline \hline
          & VER J1907$+$062 & PSR J1907$+$0602 \\
    \hline
    Configuration & D & D \\
    Band & L & C \\
    Date & 2018 Sep 3 & 2018 Sep 2\\
    Integration time & 128 min & 66 min \\
    \hline
    \end{tabular}
    \label{tab:Radio_observaciones}
\end{table}

\begin{figure*}
\centering
\includegraphics[width=\textwidth]{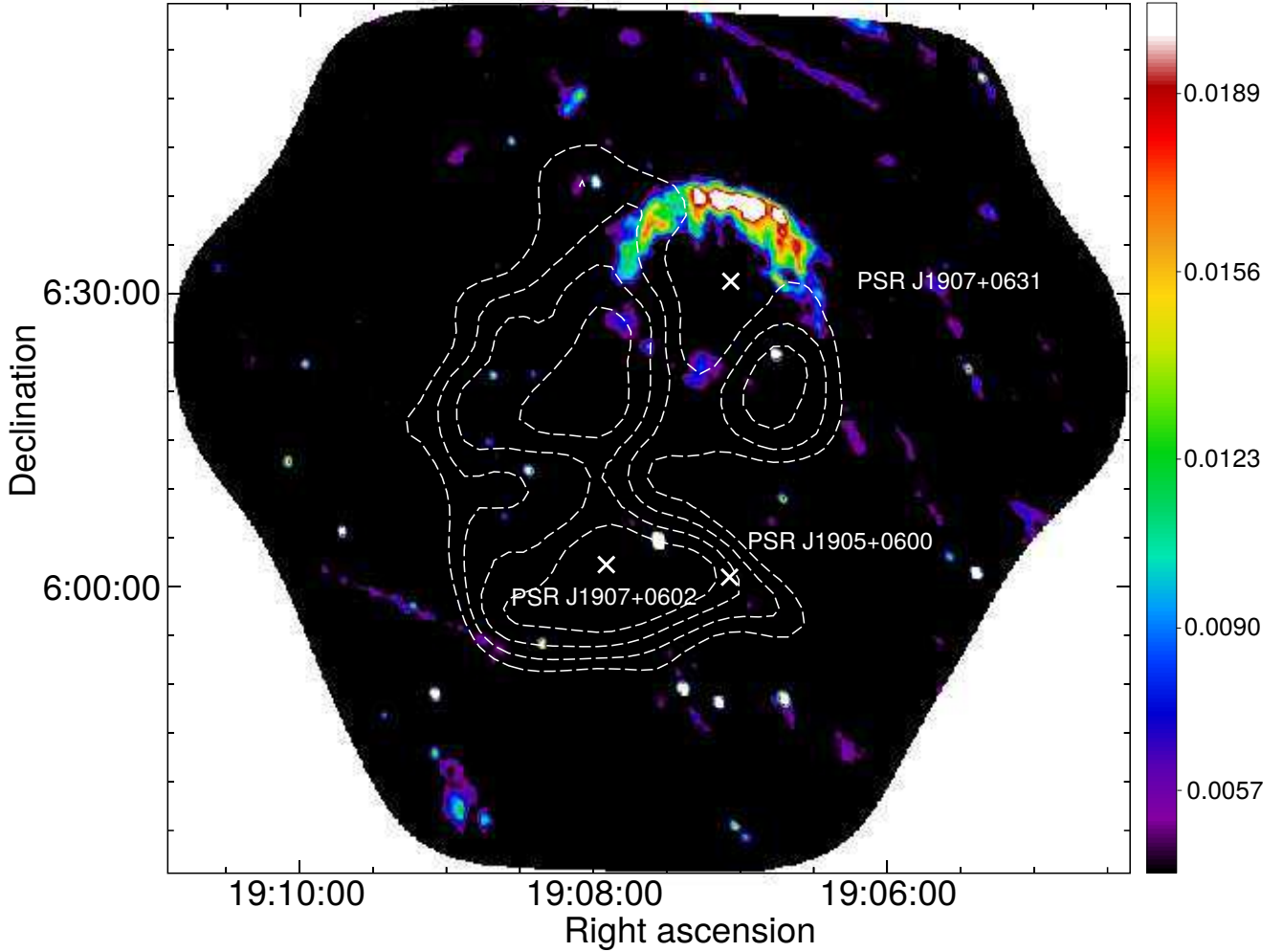}
\caption{Radio continuum image at 1.5 GHz covering the whole extension of the TeV source VER J1907$+$062. This image was obtained from a combination of 12 different pointings observed with the VLA in the D configuration. The angular resolution is 
$51^{\prime\prime}.1 \times 39^{\prime\prime}.5$, PA $= 21^\circ.9$ and the rms noise is 1 mJy beam$^{-1}$. The color scale to the right of the image is expressed in Jy beam$^{-1}$. 
The cross signs indicate the position of the pulsars PSR J1907$+$0631, PSR J1907$+$0602, and PSR J1905$+$0600. The dotted contours represent the TeV emission from VERITAS extracted from \citet{aliu14}.}
\label{Radio_hess}
\end{figure*}

\subsection{The surrounding medium}
The properties of the ISM around SNR G40.5$-$0.5 were investigated using the neutral hydrogen (HI) data extracted from the Very Large Array Galactic Plane Survey (VGPS, \citealt{stil06}), which
maps the HI 21 cm line emission with angular and spectral resolutions
of $1^\prime$~and 1.3 km s$^{-1}$, respectively.
Molecular line emission was extracted from the FOREST Unbiased Galactic
Plane Imaging\footnote{Retrieved from the JVO portal (http://jvo.nao.ac.jp/portal/) operated by ADC/NAOJ.} (FUGIN) survey, 
which maps the 1st and 3rd Galactic quadrants in the $^{12}$CO (J=1--0) using the multi-beam FOREST receiver installed on the Nobeyama 45-m telescope \citep{umemoto17}. The angular and velocity resolutions are 20$^{\prime\prime}$ and $0.65$ km s$^{-1}$, respectively.

\section{Results}

\subsection{The new VLA images} 
\label{radio_img}

The new VLA image at 1.5 GHz containing the whole extension of the TeV source VER J1907$+$062 is shown in Fig. \ref{Radio_hess}. A number of compact sources can be seen in the field together with the SNR G40.5$-$0.5. A spurious instrumental artifact visible toward the N of the field is due to the presence of the bright HII region G041.096$-$00.213 that lies outside the observed area.

The new radio data delineate the shell of the southern hemisphere of the remnant, almost lost in the continuum image of the
VGPS, thus completing a circular shell with the northern hemisphere much brighter. The recently detected PSR J1906$+$0631 lies in projection of the sky close to the center of the remnant. The enhanced radio emission at the position of the pulsar, proposed to be a probable PWN by \citet{lyne17}, is not detected with our deeper observations. We suggest that this feature, which is seen in the VGPS data, could be caused by an interference effect produced by the strong point source located at R.A. $=19^{\rm{h}}$06$^{\rm{m}}$45$^{\rm{s}}$.7, Dec. $=6^\circ 23^\prime 50^{\prime\prime}$ (J2000). On the other hand, the new radio image does not show any evidence of extended emission in coincidence with PSR J1907$+$0602 down to the noise level of our observations.

\begin{figure}
\centering
\includegraphics[width=\linewidth]{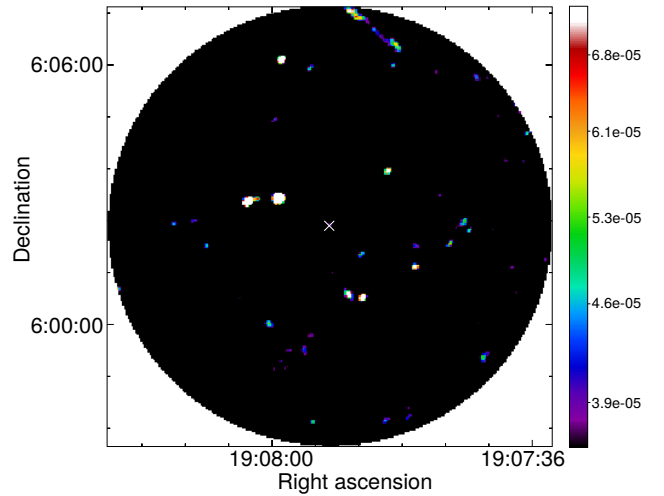}
\caption{Radio continuum image at 6 GHz of the region surrounding the position of the pulsar
PSR J1907$+$0602 (marked with a cross) and constructed using the VLA in its D array.
The final beam size, is $12^{\prime\prime}.2 \times 8^{\prime\prime}.6$ while the rms noise level is 10 $\mu$Jy beam$^{-1}$.
The color scale is expressed in Jy beam$^{-1}$. }
\label{fig:Radio_pulsar}
\end{figure}

Fig. \ref{fig:Radio_pulsar} displays the new VLA radio image at 6 GHz centered at the position of PSR J1907$+$0602 
(R.A. $=19^{\rm{h}}07^{\rm{m}}54^{\rm{s}}.73$, Dec. $=6^\circ 02^\prime 16^{\prime\prime}.9$ (J2000)). Also in this case, our new high-sensitivity radio observations show no evidence of extended nor point-like emission toward the pulsar. This result does not allow us to dismiss the suggestion made by \citet {pandel12} that the non-thermal X-ray emission around the pulsar may be a PWN. There are numerous
examples of confirmed X-ray PWN with no detected counterpart in the radio band
(see, for example, the SNRcat\footnote{Current catalog available at \\ http://www.physics.umanitoba.ca/snr/SNRcat/} of \citealt{ferrand12}).

Several deep sensitivity and high resolution radio studies searching for radio PWNe were performed with negative results
(e.g., \citealt{gaensler00,giacani14,castelletti16,sushch17}).
 In these cases, the lack of detectable radio emission has been explained by the presence of either pulsars with high magnetic field that inhibit the production of synchrotron radiation at longer wavelengths, or the presence of young and energetic pulsars residing in very low ambient density media 
($\sim$ 0.003 cm$^{-3}$), which results in severe adiabatic losses and consequently under-luminous radio PWNe.
Neither of these scenarios seem to apply for PSR J1907$+$0602. 
On one hand, its magnetic field is about 3$\times 10^{12}$ G, consistent with most of young pulsars associated with SNRs \citep{gaensler06}. 
On the other hand, to investigate the density in the environs of PSR J1907$+$0602, first we
converted from the pulsar's distance of 3.2 kpc to a velocity of $\sim 57$ km s$^{-1}$, 
using the universal rotation 
curve of \citet{persic96} with the parameters of \citet{reid14} (Galacto-centric 
distance $R_0 = 8.34$ kpc and rotation velocity of the Sun $\Theta = 241$ km s$^{-1}$) (all velocities hereafter 
are referred to the local standard of rest). 
To estimate the atomic density of the ISM where the pulsar lies, we must take into account that the DM 
does not precisely matches the kinematical distance, i.e. there are errors involved when converting from 
DM $\sim 82$ pc cm$^{-3}$ to a distance of $\sim 3.2$ kpc and hence to the kinematical velocity of $\sim 57$ km s$^{-1}$.     
In order to consider such uncertainties, we analyzed the atomic gas distribution in the velocity interval $57\pm 10$ km s$^{-1}$ 
and performed 4 integration of the HI emission into velocity intervals of $\sim 4$ km s$^{-1}$, namely $\sim [47.2 - 51.3 ]$,
$\sim [52.1 - 56.2]$, $\sim [57.1 - 61.2]$ and $\sim [62.0 - 67.0]$ km s$^{-1}$.
We estimated the HI column density $N_{HI} [cm^{-2}]=1.82 \times 10^{18} \int T_B dv$, where $T_B$ is the 
brightness temperature in K and the integration was performed in the aforementioned intervals. We assume that the emission comes
from spherical regions with a radius of $\sim 0^\circ.2$ located at distances of $\sim 2.8$, 3.1, 3.4 and 3.7 kpc, as derived
from the central velocity of each integration interval. We obtained densities between 
$\sim 5$ cm$^{-3}$ to $\sim 15$ cm$^{-3}$. Even after considering the uncertainties in the densities, which can 
be $\sim 40\%$ and are caused by the error in distance, the choice of the background level and in the integration boundaries, 
the obtained values rule out the possibility of an expansion in an area of very low density.     
 
As proposed by \citet{frail97} and 
\citet{gaensler00} these pulsars probably would be less efficient in producing radio nebulae because of the injection spectrum of particles 
in the pulsar wind has shifted to higher energies. Regarding the other pulsar, PSR J1907$+$0631, its high surface magnetic field of 
$1.2 \times 10^{13}$ G could inhibit the production of synchrotron radiation in the radio band, 
as in the case of the $\gamma$-ray pulsar RX J0007.0$+$7303 \citep{giacani14}. 

\subsection{The molecular environment of G40.5$-$0.5}
\label{molec}

The determination of the distance to the SNR G40.5$-$0.5 is important to analyze the connection with the TeV source. The only 
estimate of the distance for this remnant was based on the $\Sigma - D$ relationship, which is a method known to have large intrinsic dispersion. A more accurate method for establish distance constraints to radio continuum sources (including SNRs) 
is based on the construction of an absorption HI spectrum.
This method has given reliable distance estimates in bright SNRs (see, for example, \citealt{rana17}). 
For faint sources, the fluctuations in the HI distribution can give false absorption features
whose signals are similar to the true absorption \citep{rana18}.

We constructed an HI absorption spectrum of G40.5$-$0.5 but did not obtained an acceptable spectrum to constraint the 
source distance. Although this is a bright SNR, at least over its northern shell, the HI absorption spectrum resulted
very noisy, probably due to the fact that the neutral gas is patchy and caused spurious absorption features.

We also analyzed the distribution of the CO gas in the environment of G40.5$-$0.5 in order to identify molecular material that shows spatial correlation with both the remnant and the $\gamma$-ray emission. If an association between the SNR and molecular features is established, it serves to  better constraint its distance  
and to explore a hadronic mechanism as a probable origin for the TeV emission. After a careful inspection of the
whole $^{12}$CO (J=1--0) data cube, we found morphological signatures of a possible association in the velocity 
range between $+56$ km s$^{-1}$ and $+75$ km s$^{-1}$. 

\begin{figure*}
\centering
\includegraphics[width=0.65\linewidth]{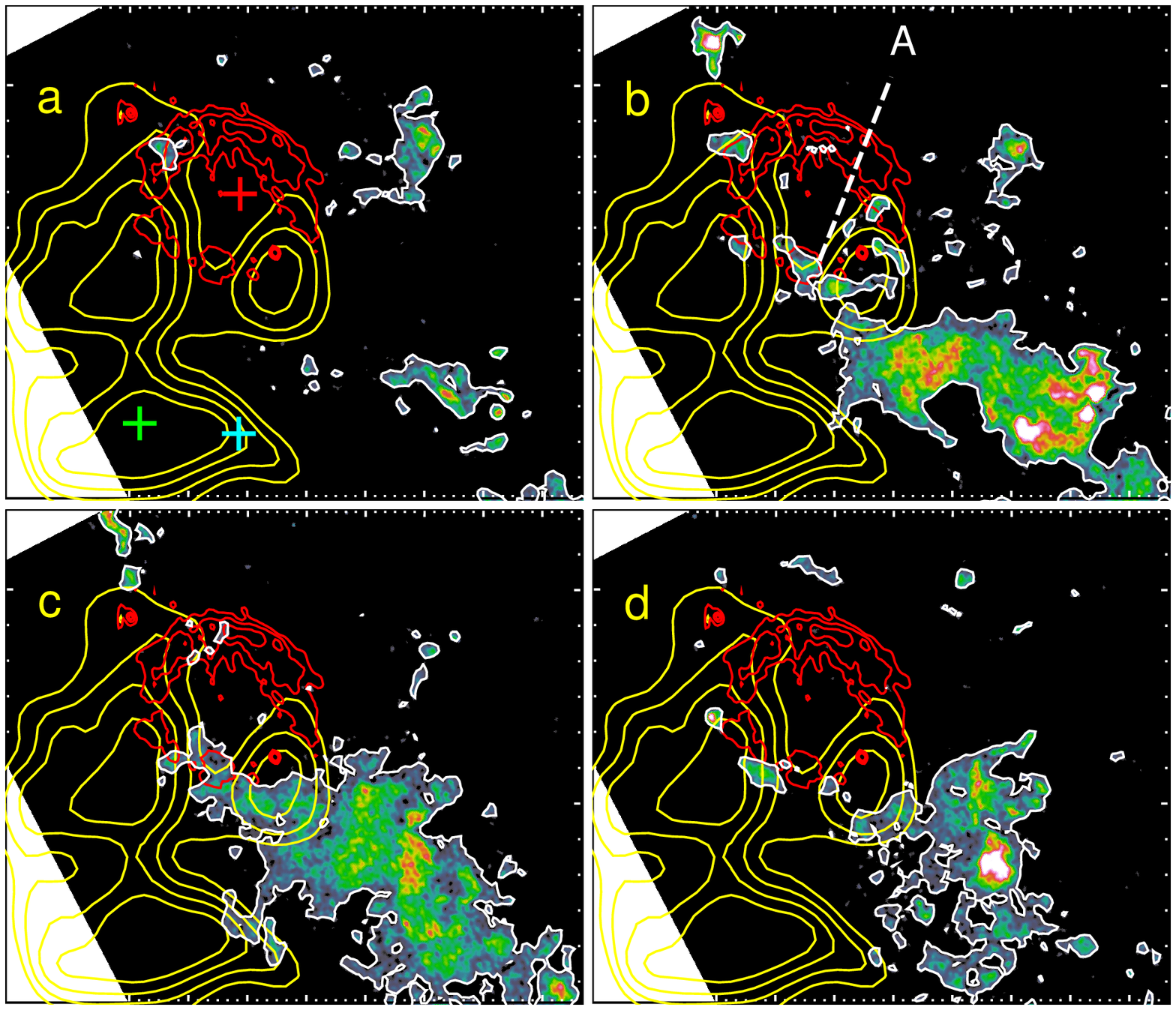}
\includegraphics[width=0.65\linewidth]{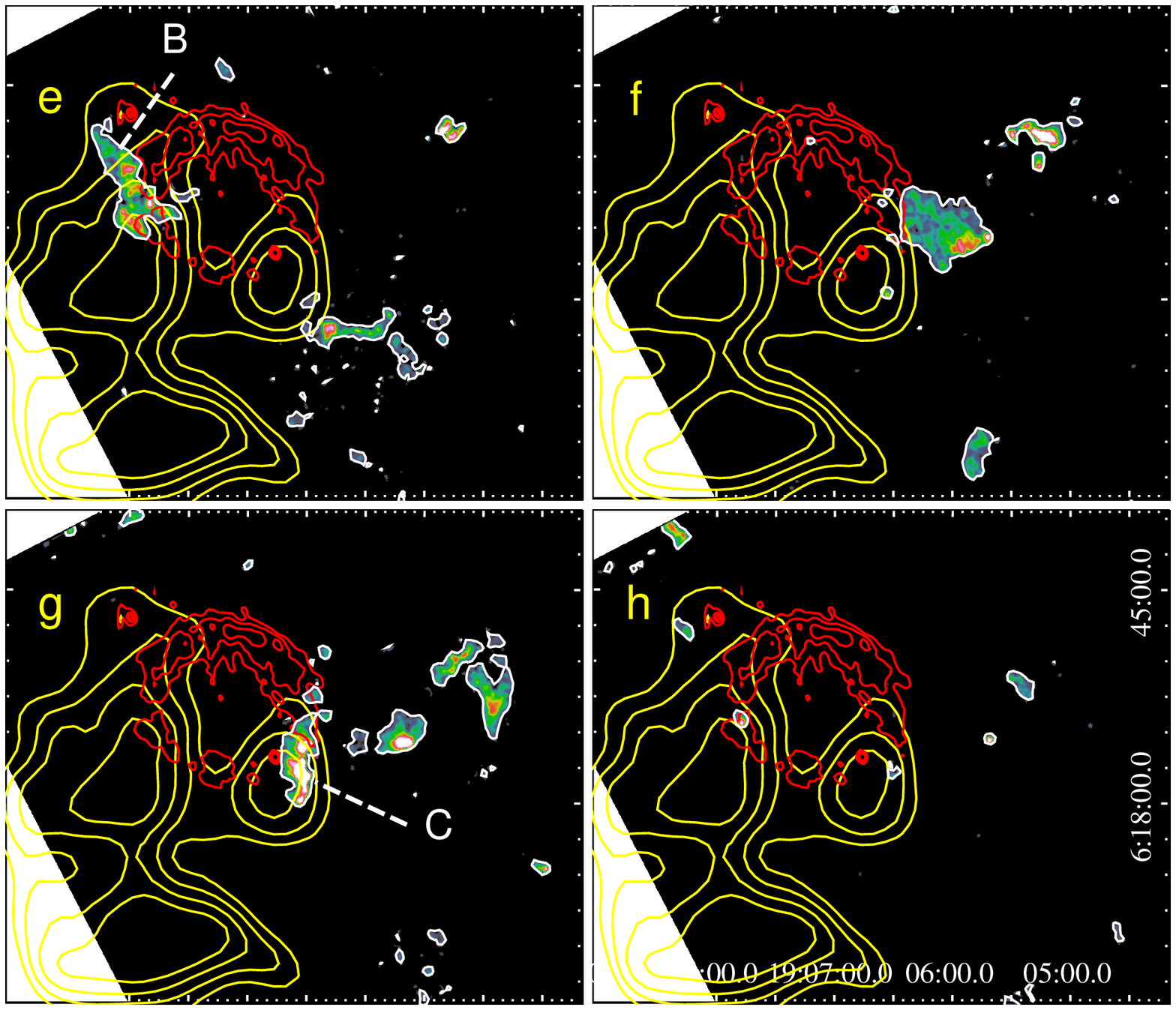}
\caption{$^{12}$CO (J=1--0) emission toward the SNR G40.5$-$0.5 integrated over $\sim 2.6$ km s$^{-1}$ velocity intervals. The velocity 
coverage of each panel is: {\it a}) $53.1 - 55.7$ km s$^{-1}$, {\it b}) $56.3 - 58.9$ km s$^{-1}$, {\it c}) $59.6 - 62.2$ km s$^{-1}$, 
{\it d}) $62.8 - 65.4$ km s$^{-1}$, {\it e}) $66.1 - 68.7$ km s$^{-1}$, {\it f}) $69.3 - 71.9$ km s$^{-1}$, 
{\it g}) $72.6 - 75.2$ km s$^{-1}$, and {\it h}) $75.8 - 78.4$ km s$^{-1}$.    
The red contours are the radio continuum emission of the SNR at 1.5 GHz and the yellow contours correspond to TeV emission from VERITAS. 
The white contour of the $^{12}$CO is the same contour level in all the panels. 
The crosses (shown only in panel {\it a}) indicate the positions of the pulsars PSR J1907$+$0602 (green),
PSR J1905$+$0600 (cyan), and PSR J1907$+$0631 (red). We label with white letters the molecular features described in the text. 
The coordinates are equatorial and are shown in panel {\it h}.} 
\label{fig:figCO_1}
\end{figure*}

Fig. \ref{fig:figCO_1} displays the $^{12}$CO distribution integrated over 
5 consecutive channels, giving a velocity coverage of $\sim 2.6$ km s$^{-1}$ for each map. The red contours delineate the radio continuum emission of the remnant at 1.5 GHz, while the yellow contours correspond to the TeV source. From this figure, we can discern several condensations of molecular gas labeled A, B, and C, bordering the edges S, E and W of G40.5$-$0.5, respectively. 
Feature A, observed in the $\sim 56-65$ km s$^{-1}$ velocity range (panels {\it b}, {\it c}, and {\it d}) 
is elongated in the NE-SW direction, and partially overlaps the southern border of the remnant and the NW maximum of the TeV source.
Feature B is seen in the $\sim 66-69$ km s$^{-1}$ range (panel {\it e}) and projected toward the eastern border of G40.5$-$0.5.
It partially overlaps the NE peak of the $\gamma$-ray emission. Finally, in the $\sim 69-75$ km s$^{-1}$ range (panels {\it f} and {\it g}), 
the molecular feature C appears overlapping the western edge of the radio shell and part of the NW maximum of the TeV emission.

We adopt $\sim 60$ km s$^{-1}$, $\sim 67$ km s$^{-1}$, and $72$ km s$^{-1}$ as the systemic velocities of molecular features A, B, and C, 
respectively. To convert from radial velocities to distances, we use again the universal rotation 
curve of \citet{persic96} with the parameters of \citet{reid14}. The first
Galactic quadrant presents distance ambiguity for positive radial velocities, so we obtain near and far distances of
$3.4$ and $9.1$ kpc for feature A, $3.9$ and $8.6$ kpc for feature B, and $4.2$ and $8.3$ kpc for feature C, respectively.   

To estimate the mass and density of the CO clouds, first 
we determine the H$_2$ column density from the integrated line flux $W(^{12}\rm{CO})$
using the empirical relation $N(\rm{H_2}) = X_{\rm{CO}} \times W(^{12}\rm{CO})$, where $X_{\rm{CO}}$ is the CO to H$_2$ conversion
factor for the $^{12}$CO (J=1--0) transition. 
We take the canonical value $X_{CO} = 2 \times 10^{20}$~cm$^{-2}$ K km s$^{-1}$ \citep{bolatto13}.
The total mass is then calculated from $M = \mu m_{H} S N(\rm{H_2})$, where $\mu$ is the mean molecular mass
($\mu=2.8$ for a relative helium abundance of 25$\%$), $m_H$ is the mass of the hydrogen atom, 
and $S$ is the area subtended by the source. The areas of the molecular features were approximated by ellipses. 
To estimate the cloud volume, we considered an ellipsoid with third axis as the mean value of the ellipse semi-axis.
For feature A, we integrated the $^{12}$CO emission in the $56 - 65$ km s$^{-1}$ velocity range.
We estimated the value $W(^{12}\rm{CO})$ from a region defined by the molecular emission superimposed to the SNR (see panel {\it b}), 
which was approximated by an ellipse of (minor~$\times$~major semi-axis) $1^{\prime}.7 \times 5^{\prime}.2$. 
For feature B, we integrated the $^{12}$CO emission between 66 and 69 km s$^{-1}$ over 
an ellipse of $1^{\prime}.9 \times 9^{\prime}.3$ 
For feature C, the integration was made in the $\sim 69-75$ km s$^{-1}$ range 
over an ellipse of $2^{\prime}.5 \times 6^{\prime}.3$.     
Since the distances are roughly compatible within the errors ($\sim 25\%$ for this method, \citealt{dubner15}), 
for the purpose of calculations, 
we adopt 3.7 and 8.7 kpc as confident near and far distances of each feature, respectively. 
The obtained masses and number densities are reported in Table \ref{tab:tabla_nube}.

\begin{table}
\caption{Physical parameters of the molecular features around the SNR G40.5$-$0.5}
\label{tab:tabla_nube}
\small
\centering
\begin{tabular}{cccc}
\hline\hline
Feature &  $D$       & $M$         &  $n$        \\
        &  (kpc)     & (M$_\odot$) &  (cm$^{-3}$)  \\
\hline
A       &  3.7       & 7200         &  660          \\
        &  8.7       & 40000        &  280      \\
\hline
B       &  3.7       & 9200       &   260 \\
        &  8.7       & 51000        & 110   \\
\hline
C       &  3.7       & 1000        & 400     \\
        &  8.7       & 55000        & 170       \\
\hline
\end{tabular}
\end{table}

It is worth noting that the far distance of the molecular features and the SNR G40.5$-$0.5 ($\sim 8.7$ kpc) is compatible, within the errors, 
with the distance to the pulsar PSR J1907$+$0631 ($\sim 8.0$ kpc) located at the center of the SNR and probably
associated with the SNR, according to \citet{lyne17}. Also, the far distance to G40.5$-$0.5 falls within the distance range 
estimated through the $\Sigma-$D relationship. Based on this fact, we suggest that $\sim 8.7$ is the most probable distance to G40.5$−$0.5.


\section{Origin of the TeV emission}
\label{TEV}

To date, the association of VER J1907$+$062 with an offset relic PWN driven by PSR J1907$+$0602 appears as the most plausible origin for the $\gamma$-ray emission. However, based mainly on the large angular extension and the morphology of the TeV emission, and its spectral behavior, the contribution of other potential particle accelerators in the field cannot be excluded as well as the possibility that the VHE emission is produced by two object superimposed along the line of sight. In what follow we explore possible scenarios that may give rise to the VHE emission.

As was mentioned in Sect. \ref{introd}, the SNR G40.5$-$0.5 lies toward the northern border of the TeV source and 
both leptonic and hadronic origin for the VHE $\gamma$-ray production will be considered. 

The acceleration of particles (both electrons and protons) at the shock front of a SNR occurs during the remnant's lifetime $t_{snr}$. 
In a leptonic scenario, the accelerated electrons interact with a low energy radiation field through the Inverse Compton (IC)
scattering, transferring their energy to the seed photons up to TeV energies.  
To produce detectable TeV photons, the electrons must have cooled via IC scattering during a time $\tau_{IC} \lesssim t_{snr}$. 
The age of G40.5$-$0.5 was estimated to be between $20-40$ kyr 
\citep{downes80}, but if the remnant 
is associated with the recently discovered PSR J1907$+$0631, G40.5$-$0.5 must 
have the pulsar's age ($\sim 11$ kyr). In order to consider all the possible ages quoted for G40.5$-$0.5, 
we adopt 11 and 40 kyr as a lower an upper limits for $t_{snr}$, respectively.    
Following \citet{kargal13}, $\tau_{IC} \sim (1+0.144 B_{\mu\rm{G}}^2)^{-1}(E_{IC}/\rm{TeV})^{-1/2}$, where $B_{\mu\rm{G}}$ is
the magnetic field in $\mu$G and $E_{IC}$ is the energy of the TeV photons.
Taking a typical $B=5$ $\mu$G, we obtain that in the lifetime of G40.5$-$0.5 only
photons with energies $E_{IC} \gtrsim 4$ TeV (for $t_{snr}=11$ kyr) and $\gtrsim 0.3$ TeV (for $t_{snr}=40$ kyr)
can be produced through the IC scattering process.
These energies are compatible with the $\gamma$-ray emission from 
VER J1907$+$062, which is detected in the $\sim 0.3 - 62$ TeV energy range \citep{abdalla18}.
We estimate the energy $E_e$ of the electrons involved in the IC mechanism from $E_{IC} \sim 4 (E_e/\rm{TeV})^2 \epsilon_{eV}$ TeV, 
where $\epsilon_{eV} = \epsilon / (1~\rm{eV})$ and $\epsilon$ is the energy of the seed photons.  
If the CMB is the main source of these low energy photons with a temperature $T\sim 3$ K ($\epsilon \sim 3 \times 10^{-4}$ eV), 
the energy of the electrons is $E_e \gtrsim 60$ TeV (for $t_{snr}=11$ kyr) and $\gtrsim 15$ TeV (for $t_{snr}=40$ kyr).

This simple calculation shows that if the VHE emission around G40.5$-$0.5 had a leptonic origin, 
there should be a population of relativistic electrons
with energies of several TeV. These high-energy electrons can interact with the ambient magnetic field and produce synchrotron 
radiation in the keV band. The energy $E_{keV}$ of the synchrotron photons
is obtained from $E_e \sim 160E_{keV}^{1/2}B_{\mu\rm{G}}^{-1/2} \rm{TeV}$ and the
corresponding synchrotron cooling time is $\tau_{sync}\sim 38B_{\mu\rm{G}}^{-3/2} E^{-1/2}_{keV}$ kyr.
If $t_{snr}=11$ kyr and taking $B=5$ $\mu$G, we get $E_{keV} \gtrsim 0.6$ keV and $\tau_{sync} \lesssim 4.4$ kyr.
If $t_{snr}=40$ kyr and taking $B=5$ $\mu$G, we get $E_{keV} \gtrsim 0.04$ keV and $\tau_{sync} \lesssim 16$ kyr.
Then, electrons accelerated at the shock front have had time enough to cool through synchrotron emission in the lifetime
of the SNR and produce X-ray photons which fall within the spectral coverage of current X-ray telescopes. However, 
{\it XMM-Newton} observations failed to detect diffuse emission in the keV band associated with the SNR \citep{pandel15}.
We note that the population of Galactic shell type SNR dominated by non-thermal X-ray emission 
is formed by a reduced number of young sources (typically $\lesssim 2$ kyr, see the aforementioned SNRcat) 
that present similar shell-like morphologies in the radio, 
X-ray and TeV bands. Representative examples of these sources are SN 1006 \citep{roth04}, G347.3$-$0.5 \citep{cassam04b}, 
Vela Jr \citep{iyudin05}, and RCW 86 \citep{vink06}.
In all of them, the VHE emission is attributed mostly to a leptonic mechanism caused by electron acceleration at the shock front
of the SNR, with the caveat that even in these sources there are hints of hadronic emission \citep{miceli14,fukui17, celli19}. Clearly, 
the SNR G40.5$-$0.5 does not match the characteristics of these sources. On the one hand, even if its age is still not firmly establish, 
all the age estimations show that it is older than 11 kyr. On the other hand, it lacks X-ray emission and the VHE emission does not match
the shell in the radio band. Thus, we disfavor a leptonic origin to explain the bulk of the VHE radiation. 


Now we analyze the hadronic mechanism, which provides an alternative explanation for VHE $\gamma$-ray emission associated with SNRs. 
In this scenario, protons accelerated at the shock front collide with target protons of the surrounding ISM 
and produce TeV photons via neutral pions decay.
Our analysis of the molecular gas around G40.5$-$0.5 revealed the presence of several clouds in good morphological correspondence
with the SNR, favoring a physical connection between them. 
We can estimate the required density matter $n_0$ in the SNR/ISM interaction region to produce
the observed $\gamma$-ray flux of VER J1907$+$062 via the hadronic mechanism. 
Following \citet{torres03}, the flux of $\gamma$-ray photons with energies greater than $E$ is given by
$F_{\gamma}(>E) = 10^{-10} f_{\Gamma} E_{TeV}^{-\Gamma+1} A$, where $f_{\Gamma}$ is a factor depending on the spectral index $\Gamma$, 
$E_{\rm{TeV}} = E / \rm{TeV}$, and $A= \theta E_{51} D_{kpc}^2 n_0$ ph cm$^{-2}$ s$^{-1}$, with $\theta$ the fraction of the SN energy that
is converted into cosmic ray energy, $E_{51}$ the mechanical energy released by
the SN in units of $10^{51}$ erg, $D_{kpc}$ is the distance in kpc, and $n_0$ the ambient density of the ISM in cm$^{-3}$.
The spectral fitting of the TeV emission from VER J1907$+$062 gives $\Gamma \sim 2.3$ (for which $f_{\Gamma} = 0.19$) and
$F_{\gamma}(> 1 \rm{TeV}) \sim 8.4 \times 10^{-12}$ ph cm$^{-2}$ s$^{-1}$ \citep{abdalla18}.
We note that this value corresponds to the TeV flux of the entire source, so this is an upper limit to the 
expected TeV flux produced by the SNR/ISM interaction. Assuming a canonical SN explosion ($E_{51}=1$) and $\theta = 0.3$,
if the SNR is located at the near distance of $\sim 3.7$ kpc, an ambient density $\gtrsim 20$ cm$^{-3}$ is sufficient
to produce the VHE $\gamma$-ray emission of VER J1907$+$062. If the SNR is at the far distance of $\sim 8.7$ kpc, the required
ambient density is $\gtrsim 120$ cm$^{-3}$. 
Taking into account that the error in the cloud's densities reported in Table \ref{tab:tabla_nube} can be of about $40\%$, we conclude that
feature A, which appears over the southern border of the remnant, is dense enough to generate the VHE emission hadronically, 
independently of whether the SNR and the CO material are either at the near or at the far distance. 
The densities of feature B and C (located over the eastern and western borders of G40.5$-$0.5, respectively), are sufficient to produce TeV photons hadronically only if the SNR and the clouds are located at the near distance of $\sim 3.7$ kpc.
Despite this, the fact that the TeV source extends far beyond the 
distribution of the target material suggests that the hadronic scenario probably is not enough to explain the entire VHE emission.   

Another possible source of TeV emission in the field is the pulsar PSR J1907$+$0631. It is seen in projection close to the center
of G40.5$-$05 and near the northern border of VER J1907$+$062 but outside the $\gamma$-ray emission region. 
The distance to the pulsar was estimated to be $\sim 8$ kpc in good agreement with the far distance of the remnant. 
Regardless of the physical connection between both objects, PSR J1907$+$0631 has a high spin-down flux
$\dot{E}/D^{2}$ of $\sim 7.8 \times 10^{33}$ erg s$^{-1}$ kpc$ ^{-2}$ for a distance $D=8$ kpc.
The TeV flux of VER J1907$+$062 in the $1-10$ TeV band is $\sim 2.3 \times 10^{32}$ erg s$^{-1}$ kpc$ ^{-2}$, requiring a conversion efficiency from rotational energy of the pulsar to $\gamma$-rays of about $3 \%$.
This is in the range of the expected conversion efficiency of known TeV PWNe ($\leq 10 \%$, \citealt{gallant07}), so 
the pulsar is energetic enough to power the entire TeV source. However, its location in a region lacking $\gamma$-ray emission and its significant offset from the centroid of VER J1907$+$062 make it an unlikely candidate as a counterpart of the VHE source. 
Even more, in the case that PSR J1907$+$062 is not associated with the SNR G40.5$-$0.5 and supposing that it was born at 
the centroid of the $\gamma$-ray emission and traveled for 11 kyr to its actual position leaving a relic TeV nebula behind, 
it would require a transverse velocity of $\sim 3800$ km s$^{-1}$ 
(for distance of 8 kpc). This is an unrealistic fast motion when compared to the velocity
of Galactic pulsars \citep{holland17,verbunt17}.   

Regarding PSR J1905$+$0600, it is an old pulsar (6 Myr) located near the SW border of VER J1907$+$062. It has a spin-down flux of about $1.6\times 10^{30}$ erg s$^{-1}$ kpc$^{-2}$ for the estimated distance of $\sim 18$ kpc, which is several orders of magnitude
lower than the TeV flux of VER J1907$+$062. Therefore it does not fulfill with the energetic requirements to power the whole VHE source.

To summarize, not a single leptonic or hadronic mechanism can fully explain the VHE from VER J1907$+$062. The most conspicuous sources of TeV radiation found in the field, namely the pulsar PSR J1907$+$0602 and the SNR G40.5$-$0.5 could account for at least a fraction of the VHE emission.
If the remnant is located at the far distance of $\sim 8.7$ kpc, its distance differs considerably from the distance of 
PSR J1907$+$0602 ($\sim 3.2$ kpc). In this case, we propose a scenario in which the TeV emission is produced by two separated
$\gamma$-ray sources. One of them is a PWN powered by PSR J1907$+$0602 and located at $\sim 3.2$ kpc. Since the pulsar lies over the 
southern maximum of the TeV source, it would be powering the $\gamma$-ray emission from the southern hemisphere of VER J1907$+$062 through 
the leptonic mechanism. The other $\gamma$-ray source, corresponding to the northern hemisphere of VER J1907$+$062, would be produced by
the interaction between the shock front of the SNR G40.5$-$0.5 with molecular material via the decay of neutral pions through an hadronic
mechanism. This VHE source would be located $\sim 8.7$ kpc away.

As suggested in Sect. \ref{molec}, the most probable distance to G40.5$-$0.5 is $\sim 8.7$ kpc, but 
we can not completely rule out the possibility that this SNR is located at the near distance of 
$\sim 3.7$ kpc. If we consider the errors involved in the estimations, 
this distance is compatible with the distance of PSR J1907$+$0602. Also in this case, VER J1907$+$062 could be
the superposition in the line of sight of two nearby, but still separated, $\gamma$-ray sources associated with 
PSR J1907$+$0602 and with the SNR G40.5$-$0.5. Alternatively, if the pulsar and the SNR are at the same distance, VER J1907$+$062
could be a single $\gamma$-ray source whose emission is powered by the combination of two mechanisms: a leptonic one, associated with
a TeV PWN driven by PSR J1907$+$0602, and a hadronic one, associated with the interaction between the SNR G40.5$-$0.5 and the surrounding
molecular gas. 

An improved statistic in the TeV band is necessary to confirm whether VER J1907$+$062 is the superposition in the line of sight 
of two distinct $\gamma$-ray sources powered by different emission mechanisms and located at different distances, 
or whether it its a single source whose VHE emission is produced by two particle accelerators (the pulsar and the SNR) located
at the same distance. The unambiguous determination of the distance of the SNR G40.5$-$0.5 could shed light
on which of the proposed scenarios is taking place.  

\section{Summary}

In this paper we present new high-quality radio images of a large region containing the extended TeV source VER J1907$+$062 at 1.5 GHz and a region toward the PSR J1907$+$0602 at 6 GHz, in both cases with data obtained using the VLA in its D configuration. In spite of the quality of our images, no nebular radio emission has been found toward PSR J1907$+$0602 (which presents hints of an X-ray PWN) 
neither toward the other two pulsars in the region, PSR J1907$+$0631 and PSR J1905$+$0600. Besides, there is no indication of the existence of a host SNR associated with the mentioned pulsars.
From the analysis of the spatial distribution of the $^{12}$CO in the vicinity of the SNR G40.5$-$0.5, we found molecular clouds
which match the eastern, southern, and western borders of the remnant and partially overlap peaks of the TeV emission from VER J1907$+$062.
The distance to these clouds (and hence to the SNR) was constraint to be either $\sim 3.7$ or $\sim 8.7$ kpc. Based on the distance to the pulsar PSR J1907+0631 and its location at the center of the SNR G50.5$-$0.5 we suggest the far distance as the most probable to the remnant.
We discuss  possible origins to explain the nature of VER J1907$+$062 and favor a scenario in which the TeV source consists of 
two separate $\gamma$-ray sources. One of them of leptonic origin driven by the pulsar 
PSR J1907$+$0602 located at $\sim 3.2$ kpc, and the other source of hadronic origin through collisions of ions accelerated at the 
shock front of the SNR G40.5$-$0.5 with the molecular gas in its surroundings at a distance $\sim 8.7$ kpc.


\section*{Acknowledgements}

The authors wish to thank the anonymous referee since his/her comments improved our manuscript. L.D. is doctoral fellow of CONICET, Argentina. 
A.P. and  E.G. are members of the {\sl Carrera del Investigador Cient\'\i fico} of CONICET, Argentina. 
This work was partially supported by Argentina grants awarded by UBA (UBACyT) and ANPCYT.
This publication makes use of data from FUGIN, FOREST Unbiased Galactic plane Imaging survey with the Nobeyama 45-m telescope, 
a legacy project in the Nobeyama 45-m radio telescope.




\bibliographystyle{mnras}
\bibliography{ref} 








\bsp	
\label{lastpage}
\end{document}